\documentclass[reprint, aps, prd, superscriptaddress,twocolumn,amsmath,amssymb,nofootinbib]{revtex4-1}
\usepackage{graphicx}
\usepackage{color}
\usepackage{epstopdf,cancel,ulem}
\usepackage{epsf,latexsym,bbm,euscript}
\usepackage{amssymb,amsmath}
\usepackage{mathtools} %used in this file for the command \coloneqq -> :=
\usepackage{times,graphics}
\usepackage{soul,xcolor}
\usepackage{epsfig}

\usepackage[colorlinks]{hyperref}

\newcommand{\be}{\begin{equation}}
\newcommand{\ee}{\end{equation}}
\newcommand{\ba}{\begin{eqnarray}}
\newcommand{\ea}{\end{eqnarray}}

\newcommand{\beq}{\begin{equation}}
\newcommand{\eeq}{\end{equation}}
\newcommand{\beqa}{\begin{eqnarray}}
\newcommand{\eeqa}{\end{eqnarray}}

\newcommand{\tcb}{\textcolor{blue}}

\newcommand{\C}[1]{{\cal{#1}}}

\def\Gdp{\Gamma_{\mathrm{DP}}}
\def\Gktm{\Gamma_{\mathrm{KTM}}}
\def\tGktm{\tilde{\Gamma}_{\mathrm{KTM}}}

\newcommand{\ket}[1]{\lvert \, #1\rangle}
\newcommand{\bra}[1]{\langle #1 \, \rvert}

\newcommand{\h}[1]{\hat#1{}}
\def\Tr{{\textrm{Tr}}}

\usepackage{comment} % comment-out piece of text
\usepackage{color} %needed for coloring of \todo etc.

\newcommand{\pau}[1]{\textcolor{magenta}{#1}}

\begin{document}
\title{Gravity is not a Pairwise Local Classical Channel}
%Experiments rebut classical channel model of gravity

\author{Natacha Altamirano}
\email{naltamirano@perimeterinstitute.ca}
\affiliation{Perimeter Institute for Theoretical Physics, 31 Caroline St. N. Waterloo Ontario, N2L 2Y5, Canada}
  \affiliation{Department of Physics and Astronomy, University of Waterloo, Waterloo, Ontario, Canada, N2L 3G1}
\author{Paulina Corona-Ugalde}
\email{pcoronau@uwaterloo.ca}
\affiliation{Institute for Quantum Computing, University of Waterloo, Waterloo, Ontario, N2L 3G1, Canada}
 \affiliation{Department of Physics and Astronomy, University of Waterloo, Waterloo, Ontario, Canada, N2L 3G1}
 \author{Robert B. Mann}
\email{rbmann@uwaterloo.ca}
\affiliation{Perimeter Institute for Theoretical Physics, 31 Caroline St. N. Waterloo Ontario, N2L 2Y5, Canada}
 \affiliation{Institute for Quantum Computing, University of Waterloo, Waterloo, Ontario, N2L 3G1, Canada}
  \affiliation{Department of Physics and Astronomy, University of Waterloo, Waterloo, Ontario, Canada, N2L 3G1}
 \author{Magdalena Zych}
\email{m.zych@uq.edu.au}
\affiliation{Centre for Engineered Quantum Systems, School of Mathematics and Physics, The University of Queensland, St Lucia, Queensland 4072, Australia}

\date{\today}

%%%% OLD ABSTRACT %%%%%%%%%%%

%It is currently believed that there is no experimental evidence on gravity-inspired modifications to quantum mechanics or decoherence models. Here we show that single-atom interference experiments achieving large spatial superpositions can rule out a particular realization of a gravitational decoherence model of Kafri, Taylor and Milburn (KTM): where gravitational interactions act pairwise between massive particles as classical channels, approximating Newtonian pair-potential at low energies. Specifically, experiments indicate that if gravity does reduce to pairwise Newtonian interactions between atoms in a non-relativistic limit, these interactions cannot be fundamentally classical.  We discuss in detail how the KTM model differs from the model of Diosi and Penrose, which is not constrained by the same data.
\begin{abstract}
It is currently believed that there is no experimental evidence on possibly quantum features of gravity or gravity-motivated modifications of quantum mechanics. Here we show that single-atom interference experiments achieving large spatial superpositions can rule out a {framework} where the Newtonian gravitational interaction is fundamentally classical in the information-theoretic sense: it cannot convey entanglement. Specifically, {in this framework} gravity acts pairwise between massive particles as classical channels, which effectively induce approximately Newtonian forces between the masses. The experiments indicate that if gravity does reduce to the pairwise Newtonian interaction between atoms at the low energies, this interaction cannot arise from the exchange of just classical information, and in principle has the capacity to create entanglement. We clarify that, contrary to current belief, the classical-channel {description} of gravity differs from the model of Diosi and Penrose, which is not constrained by the same data. %\red{Our results raise the question what is the meaning of the classically of gravity implied by the latter model.}

\end{abstract}
% total # of words in the entire document @ the beginning: 5812
% need to reduce by ~210
%v4 total: 5412 (2634)
%current total: 5469 (2691)

%\preprint{DAMTP-2011-80}
% 04.50.-h  Higher-dimensional gravity and other theories of gravity
% 04.50.Gh  Higher-dimensional black holes, black strings, and related objects
% 04.70.Bw  Classical black holes
% 04.20.Jb  Exact solutions
% 05.70.Ce Thermodynamic functions and equations of state
% 04.70.-s Physics of black holes
%\preprint{pi-stronggrv-291}

\maketitle

%\section{Introduction} 

There is  an overwhelming experimental evidence  that  properties of local physical systems are incompatible with a  {fully local} classical description \cite{Aspect:1981,Giustina:2013, 
Hensen:2015loophole,Giustina:2015.Significant,Shalm:2015.Strong, kirchmair2009state,  lapkiewicz2011experimental,hu2016experimental}.  
Nevertheless, the possibility that gravity remains classical at a fundamental level is considered viable or even necessary \cite{ref:Karolyhazy1966, hall2005interacting, 
PhysRevD.78.064051, hall2016ensembles, ref:Diosi1989, Diosi2007NJP, Tilloy:2015zya, Penrose:1996, Penrose2014, jacobson1995thermodynamics}, with a range of arguments invoked to support such a position: absence of direct observations of  quantum gravitational phenomena \cite{Page:1981aj}, anticipated pernicious  tensions  between the foundational principles of quantum theory and general relativity \cite{Penrose:1996, Penrose2014, Isham:1993canonical} (see e.g.~\cite{Eppley:1977,marletto2017we} for different views), and lack of a complete framework for quantum gravity \cite{Kiefer:2014sfr}. 

%Our results, while not proving that gravity can entangle massive particles, suggest that fundamentally gravity is not classical, providing yet another strong argument to seek for .experiments testing gravitational entanglement
%this has recently beed applied to gravity: 

From an information-theoretic perspective, classicality of an interaction is defined as the inability of the resulting channel to increase entanglement. {Thus, in order to verify whether gravity is a quantum or a classical entity it has been proposed to test its entangling capacity using a pair of masses in two close-by interferometers~\cite{marletto2017witness, marletto2017entanglement, bose2017spin}.

Here we take a different approach and explore consequences of the assumption that gravity is fundamentally classical in the information-theoretic sense, and is incapable of creating entanglement.} Since a unitary interaction in general does increase entanglement,  an interaction with a known unitary part must be accompanied by decoherence in order for the resulting channel to be entanglement non-increasing -- a model-independent result shows that this decoherence must be at least twice the interaction strength \cite{2013arXiv1311.4558K, Altamirano2016knc} %for the resulting channel to be entanglement non-increasing 
(see also~\cite{WisemanMilburn:Book:2010, jacobs2014BookQuantum} for a broader context of effective dynamics in a classical stochastic environment).

The presence of the unitary Newtonian term in the Schr{\"o}dinger equation is experimentally well established  \cite{COW:1975, nesvizhevsky2002quantum, rosi2014precision, PhysRevLett.114.013001, asenbaum2016phase}. Therefore, for gravity to be a fundamentally classical channel %connecting pairs of masses 
the unitary Newtonian term must be accompanied by certain minimal decoherence -- first shown in a series of works by Kafri, Taylor and Milburn (KTM) \cite{2013arXiv1311.4558K, Kafri:2014zsa, Kafri:2015iha}. The significance of the KTM approach is that {it provides a broad framework for understanding how to describe gravitational interactions in an information-theoretic manner, and their} lower bound on decoherence distinguishes theories where low-energy particles can or cannot develop entanglement through the Newtonian interaction.

Here we show that this information-theoretic notion of classicality of gravity is incompatible with the results of recent atom interference experiments \cite{kovachy2015quantum, SugarbakerThesis}, %where the total loss of coherence was much smaller than the KTM bound required for gravity to remain classical. 
{heavily constraining the possibility that gravity acts as pairwise classical channels effectively inducing Newtonian force at low energies}. %Current experiments,  while not proving that gravity does entangle massive particles, %they already  that gravity cannot be fundamentally classical 
% already provide a strong argument for the physical relevance of gravity-mediated entanglement, 
{While current experiments do not directly prove that gravity does entangle massive particles}, %they provide a strong argument for the physical relevance of gravity-mediated entanglement, 
{they nevertheless constrain the same model that would be tested in experiments proposed in refs}
\cite{marletto2017witness, marletto2017entanglement, bose2017spin}.
%{refuting the possibility that gravity acts as pairwise {\it{classical}} channels yielding Newtonian potential at low energies.} 
Furthermore, we show that decoherence resulting from the KTM approach is conceptually and quantitatively different from decoherence in the Diosi-Penrose (DP) and related models \cite{%ref:Karolyhazy1966, 
ref:Diosi1989, Diosi2007NJP, Penrose:1996, %ref:Anastopoulos2013, 
Bassi:2012bg, Tilloy:2015zya}, not refuted by the same  data. %While DP model is widely considered to allow for gravity to remain classical, our results rise the 

\section{Effective gravity from local, classical channels} %continuous measurements and feedback\; } 

{The KTM framework  is an  application of pairwise continuous measurement with feedback \cite{WisemanMilburn:Book:2010, jacobs2014BookQuantum} to gravitational interactions.}  
 It can also be obtained from a quantum collisional model, where the systems interact with a Markovian environment in a suitably chosen parameter regime \cite{Altamirano2016knc}.  
Below we  summarize key aspects of {this approach}~\cite{Kafri:2014zsa} (see also Appendix \ref{app:KTM}).

Consider a pair of particles  interacting with a set of ancillae (environment). {The assumptions are}: a) particles interact with the ancillae but not with each other; b) the interactions are local and any information transmitted through the ancillae is classical  c) the unitary part of the channel reduces to the standard Newtonian pair-potential at low energies. %Assumptions a) and b) mean that gravity acts as an LOCC channel (Local Operations and Classical Communication) connecting pairs of masses. %Thus, the interactions are local in the position of the particles and can in general be interpreted as a  continuous measurement of each particle's position with strength $D$ and a feedback with strength $K$ -- chosen so as to give rise to the Newtonian potential. %The measurement outcomes determine the magnitude of the feedback, which is a shift of the particle position. 
 The ancilla can here be {regarded} as gravitational degrees of freedom and they facilitate measurement-and-feedback scenario,  equivalent to averaging over definite but unknown measurement outcomes and correspondingly applied local feedback. {The framework thus defines} a {Local Operations and Classical Communication 
(LOCC) channel} \cite{NielsenBook2000} between the pair of masses.
The resulting dynamics of two particles along the radial, $x$, direction is obtained by tracing over the ancilla and reads
\beq
 \dot{\rho}_{12}\!=\!-\frac{i}{\hbar}[\h H_0+V_0, \rho_{12}] 
 -\bigg(\frac{1}{4D}+\frac{K^2D}{4\hbar^2}\bigg)\!\sum_{i=1}^2[\h x_i, [\h x_i,\rho_{12}]], 
 \label{twopart_master_1}
 \eeq
where $\dot{\rho}_{12}$ is the state of both particles and  $\hat x_1$,  $\hat x_2$ is the displacement from the initial position of the respective particle. The effective unitary interaction %in eq.~\eqref{twopart_master_1} 
$V_0=K\hat x_1\hat x_2$+ (local terms in $\hat x_1$, $\hat x_2$), for $K:=2\frac{Gm_1m_2}{d^3}$  approximates the Newtonian potential between masses $m_1$, $m_1$ at a distance $d+x_1+x_2$; i.e.~$V_0\approx -G\frac{m_1m_2}{|d+x_1+x_2|}$ up to second order in $x_i$, $i=1,2$. It is accompanied by non-unitary terms, given by the double commutators, describing decoherence in the position basis: 
For each particle, the magnitude of its off-diagonal elements $x_i$, $x^\prime_i$ decays at a rate 
$\Gktm=\left(\frac{1}{4D}+\frac{K^2D}{4\hbar^2}\right)\Delta x^2$, where $\Delta x =|x_i - x^\prime_i|$ is the ``superposition size''
of the $i$-th particle. Importantly, $\Gktm$ has a non-vanishing lower bound $\propto\frac{K}{2\hbar}$. %The KTM model is thus essentially parameter-free. 
 The decoherence rate of each particle is thus fully characterised by the gradient of the Newtonian force between the masses, $\frac{K}{2}=\frac{Gm_1m_2}{d^3}$, and by the superposition size $\Delta x$:
\beq
\label{gamma_ktm_min}
\Gktm^{min}=\frac{K}{2\hbar}\Delta x^2.
\eeq
The effective interaction is necessarily accompanied by decoherence of at least the same strength since LOCC channels are entanglement non-increasing \cite{NielsenBook2000} (see also ref.~\cite{Altamirano2016knc} for a discussion in this specific context).
If the decoherence rate is smaller than $\Gktm^{min}$, the unitary term $V_0$ can increase entanglement between the particles \cite{Kafri:2014zsa} -- this is independent of the specific model for the channel or the ancilla. Any dynamical theory of gravity giving rise to the same unitary term as in eq.~\eqref{twopart_master_1} but with smaller decoherence, can in principle generate entanglement and is therefore not fundamentally classical, (not compatible with an LOCC channel).

%%%%%%%%%%%%%%%%%%%%%%%%%%%%%%%%%%%%%%%%%%%%%%%%
%%%%%%%%%%%%%%%%%%%%%%%%%%%%%%%%%%%%%%%%%%%%%%%%

\subsection{Composite systems}  

We shall apply the KTM approach to a pair of systems comprising an atom in an interferometer and the Earth. We first demonstrate that upon extending the KTM model to macroscopic systems %the key aspect of the KTM model, 
the lower bound on decoherence \eqref{gamma_ktm_min} remains unchanged up to a factor related to the geometry of the bodies 
(see Appendix \ref{app:composite_grav}). 

 Let  $s_1$, $s_2$ be rigid bodies with total masses $M_1$, $M_2$, comprising $N_1$, $N_2$ elementary constituents, respectively, with masses $m_i$, $i=1,..., N_1+N_2$.  We take  the
minimum of the decoherence rates, as in eq.~\eqref{gamma_ktm_min}, for each pair, whose evolution is
described by eq.~\eqref{twopart_master_1}. %Since we focus on earth-based atom interferometrce experiments, 
We consider $s_1$ to be a test mass, in a superposition of different radial distances from  the body $s_2$ which describes all the remaining matter (Earth, $\sim 500$ kg of tungsten \cite{rosi2014precision}, etc) and is thus considered initially well localised.
The resulting dynamics of the {centre-of-mass (CM)} of $s_1$, in the radial direction, is described by
\begin{equation}
\label{composite_master}
\begin{aligned}
\dot{\rho}_{s_1}=-\frac{i}{\hbar}[\h H_0+ V, \rho_{s_{1}}] - \mathcal{D}_{min}[\h r_1, [\h r_1,\rho_{s_1}]],
\end{aligned}
\end{equation}
where $V\approx -G\frac{M_1M_2}{|d+ r_1+r_2|}$, with $r_k$ the displacement of the CM of $s_k$ and  
\begin{equation}
\label{decoh_min}
\mathcal{D}_{min}:=\frac{1}{2\hbar}\left(\sum_{i\neq j\in s_1} |K_{ij}|+\sum_{i\in s_1}\sum_{j\in s_2}|K_{ij}|\right),
\end{equation}
 where $K_{ij}$ 
 is the Newtonian force gradient between the masses $m_i$, $ m_j$ in three dimensions. The non-unitary term is simply the sum of pairwise contributions from all constituents of the bodies. 
The corresponding minimal decoherence rate reads
\beq
\label{gamma_multipart_min}
\tGktm^{min}=\mathcal{D}_{min}\Delta x^2.
\eeq
The sum of the unitary contributions approximates Newtonian interactions between all constituents -- which is the gravitational potential energy between two point masses $M_1, M_2$. 
However, the decoherence rate  \eqref{decoh_min} in general differs from that for two elementary masses: first,
 it contains terms connecting constituents of $s_1$ (first sum), and $s_1$ with $s_2$ (second sum).
 Second, for non-convex bodies the rate~\eqref{gamma_multipart_min} might be smaller than the rate given by the original {model,} % proposed by KTM
eq.~\eqref{gamma_ktm_min} applied to the CMs of $s_1, s_2$ (e.g.~when $s_2$ is a spherical shell of matter with $s_1$  at the centre). 
For an elementary test mass $m$ near the surface of a homogeneous ball of mass $M$ and radius $R$, the KTM decoherence rate is at least as large as (see Appendix \ref{app:composite_grav})  
\beq
\label{gamma_multi}
{\Gktm^{min}\geq}  \tGktm^{C} = C\frac{GMm}{\hbar R^3}\Delta x^2,
\eeq
with $C \simeq 0.47$ for this particular geometry. For $C=1$ the above reduces to the original KTM model applied directly to the CMs of the two systems.

Note that in general one cannot here approximate the mass distributions to be continuous, since contributions from the body's own constituents diverge and an explicit definition of the \textit{fundamental} constituents is needed.
We propose that these should be the smallest constituents between which the binding energy contribution to the total mass can be neglected (since the total mass of the system is here the sum of the masses of its constituents). We hereafter consider atoms as such fundamental constituents. 

\section{KTM vs Atomic Fountains} 

{We can now confront the KTM proposal} against two interferometric tests with atoms that use large momentum transfer (LMT) \cite{kovachy2015quantum, SugarbakerThesis}. We treat the interfering atom as a test mass $s_1$ and Earth as the massive ball $s_2$ in eq.~\eqref{gamma_multi}.  
\begin{figure*}[ht!]
 \includegraphics[width=0.87\textwidth]{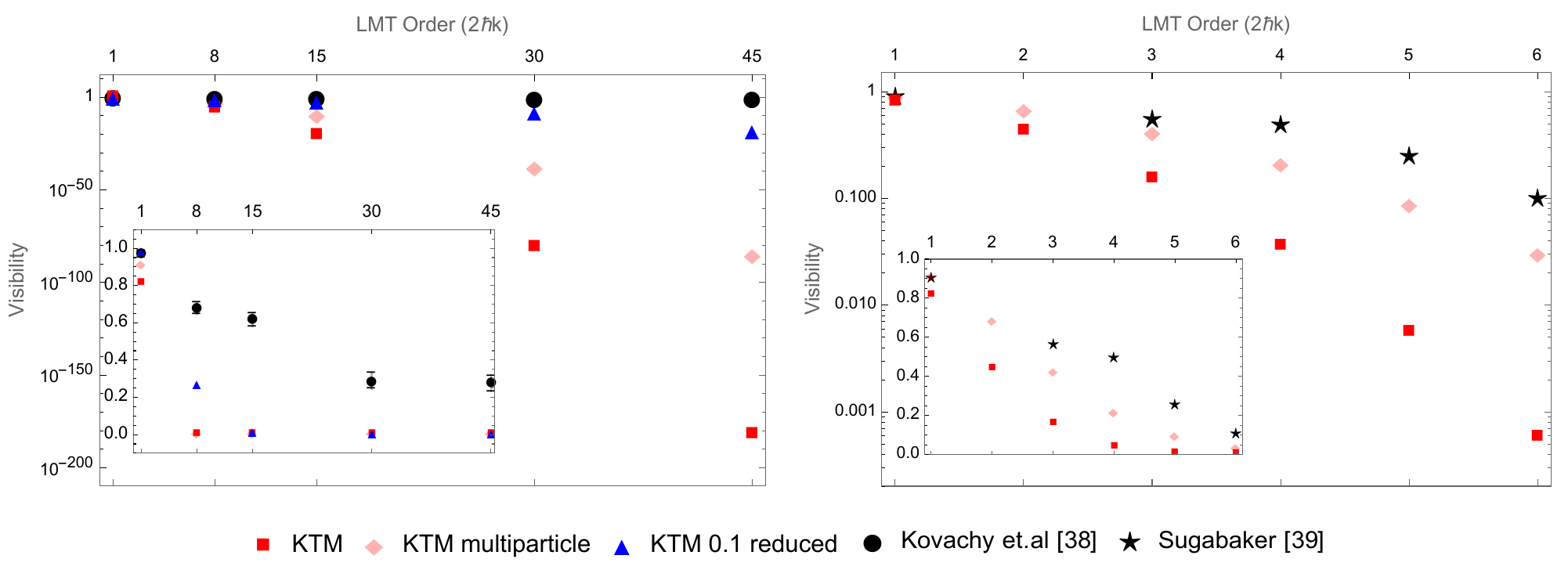}% \includegraphics[width=.4\textwidth]{plot_sugarbaker.pdf}\qquad\qquad\includegraphics[width=0.9\textwidth]{legends1.pdf}
\caption{
Logarithm of the interferometric visibility as a function of the superposition size (LMT order) reported in ref.~\cite{kovachy2015quantum} (black dots, left panel --  including the reported error bars) and in ref.~\cite{SugarbakerThesis} (black stars, right panel) vs the prediction of the KTM model eq.~\eqref{visib_atomf} for $C=1$ (red squares), its multi particle correction $C=0.47$ (pink diamonds), reduced KTM correction $C=0.1$ (blue triangles). (The insets represent the same data in a linear scale.) {Both experiments used  $^{87}$Rb; the reported superpositions were up to $8.2$ cm in \cite{SugarbakerThesis} and $54$cm in \cite{kovachy2015quantum}. Both experiments contravene the KTM model (including its multi-particle formulation).}}
\label{fig1}
\end{figure*}

In LMT interferometers a sequence of $N$ $\frac \pi 2$ laser pulses implements a beam splitter, preparing the atoms in a superposition of wave packets with momentum difference $2N\hbar k$ in the vertical direction, where $k$ is the laser  wave-number.
For time $T$ the wave packets propagate freely and thus spatially separate; then a sequence  of $\pi$-pulses exchanges their momenta and at time  $2T$ the wave packets interfere at a final beam splitter ($N$ $\frac \pi 2$-pulses). For an atom of mass $m$ the vertical separation between the wave packets is $\Delta x(t)=2N\hbar k  t/m$ for $t<T$, it then symmetrically decreases until $t=2T$.  Eq.~(\ref{composite_master}) entails that the magnitude of the off-diagonal elements of the atom $V(t):=| \bra{r_1}\rho_{s_1}\ket{r_2}(t)|$ at the end of the interferometric sequence reads $V(2T)=| \bra{r_1}\rho_{s_1}\ket{r_2}(0)|e^{-\int_0^{2T} dt\mathcal{D}_{min} \Delta x^2(t)}$. Since $V(2T)$ describes the visibility of the interference pattern attainable in the experiment, the maximal visibility allowed by {the classical channel framework} for atom fountains on Earth is estimated as $e^{-2\int_0^T dt \,\tGktm^{C}}$, which for the above $\Delta x(t)$ reads:
\be\label{visib_atomf}
V_{KTM}^{max}=e^{{-\frac{2}{3}C\frac{ G\hbar M_\oplus}{m R_\oplus^3}(2Nk)^2T^3}}.
\ee
In Fig.~\ref{fig1} we compare this prediction for $C=1$ (the original KTM model), $C=0.47$ (our multi particle correction), and $C=0.1$ (arbitrary down-scaling of the decoherence rate) against measured visibilities \cite{kovachy2015quantum} and \cite{SugarbakerThesis}
(noting at this point the controversy \cite{stamper2016verifying} regarding ref.~\cite{kovachy2015quantum}). The values of the relevant parameters are $M_\oplus=6\cdot10^{24}$ kg; $R_\oplus=6\cdot10^3$ km, $\frac{\hbar k}{m}=5.8$ $\frac{\textrm{mm}}{\textrm{s}}$,  $m=1.4\cdot 10^{-25}$ kg ($^{87}$Rb);  $T=1.15$ s in Ref.~\cite{SugarbakerThesis} and $T=1.04$ s in Ref.~\cite{kovachy2015quantum}. 

Both the original KTM model ($C=1$) and the multi particle correction ($C=0.47$) predict maximal visibilities that are well below the measured ones.  Taking into account finite duration of light-atom interactions would introduce a correction to the path separation. In the insets of Fig.~\ref{fig1} we show that even if this resulted in a reduction of the decoherence rate to $C=0.1$, the resulting visibilities would still be smaller than the measured ones by factors ranging from $\sim2.5$ to $\sim10^{18}$. 

\section{Comparison to the Diosi-Penrose model}
\renewcommand{\arraystretch}{1.2}
\begin{table*}[ht!]
\caption{Comparison of decoherence times $1/\Gdp$ and $1/\Gktm^{min}$ predicted by DP  and  KTM models  for matter-wave interference experiments \cite{rosi2014precision, kovachy2015quantum,  ref:Eibenberger2013, sclafani2013quantum}. For all tests $1/\Gktm^{min}$ contains the contribution from Earth $M_\oplus\sim6\times 10^{24}$ kg,  $R_\oplus\sim 6\times 10^3$ km,  and for experiment \cite{rosi2014precision} additionally from $24$ bars of tungsten\footnote{Tungsten bars were used to induce controllable gravity gradient between two  atom interferometers operating as gravity gradiometer.}  
each with mass $\sim21.5$ kg, and 6 of each at approximate distances of {$107.6$} mm,  {$177.6$} mm,  {$279.5$} mm and {$313.1$} mm from the atoms. For $\Gdp$ we took $\delta=10^{-15}$. 
Labels deonte: $m$ -- mass of the interfering particles; $M$ -- ``source'' mass/masses; $d$ -- distance between the source and the test mass (both relevant only for $\Gktm$); $\Delta x$ -- superposition size. For simplicity we treat all test masses as single particles. %The KTM decoherence time $\sim10^{-3}$s  (first row) is much shorter than $2$ s  timescale of the experiment \cite{kovachy2015quantum}, which we analyse in detail further in this work.
\hspace*{\fill}}
\centering
\begin{tabular}{||c l |c | c| c | c | c | c  | c ||}
\toprule
   &  Experiment   &     m [Kg]    &    M [Kg]   & $d$ [m]  & $\Delta x$ [m]  &   $1/\Gdp$ [s] &   $1/{\Gktm^{min}}$ [s] \\ \hline
  
  &   10 m  atomic fountain with $^{87}$Rb \cite{kovachy2015quantum}   &     $1.4\times 10^{-25}$    & $M_\oplus$   & $R_\oplus$  & $0.54$  &   $3\times 10^{10}$ &$2\times 10^{-3}$ \\ \hline
 
   & two atomic fountains with $^{87}$Rb  \cite{rosi2014precision} &     $1.4\times 10^{-25}$    &    $M_\oplus$   &  $R_\oplus$ & {$1.86\times10^{-3}$}  &   {$3\times 10^{10}$} & {$2\times 10^{1}$} \\
 
   & (operating as gravity-gradiometer)   &        &   {$ 4 \times 129$ }  & {$0.11, 0.18, 0.28, 0.31 $}  &   &    & \\ \hline
 
  &  large-molecule interferometry \cite{ref:Eibenberger2013}   &    $1.6\times 10^{-23}$   &  $M_\oplus$   &  $R_\oplus$   & $2.7\times10^{-7}$   & $3\times 10^6$&   $6\times 10^7$ \\ \hline
 & PcH$_2$ diffraction on alga skeleton \cite{sclafani2013quantum}   &    $8.2\times 10^{-25}$   &  $M_\oplus$   &  $R_\oplus$   & $2\times10^{-7}$   & $1\times 10^{9}$&   $2\times 10^9$ \\ \hline\hline
   %\bottomrule	   
\end{tabular}
\label{testtable}
\end{table*}
The experiments refuting the classical-channel description of gravity do not constrain the DP model; see Table \ref{testtable}. {As the KTM approach and the DP model are thought to give equivalent decoherence rates \cite{Kafri:2014zsa,Tilloy:2015zya},  we discuss below their key differences.} In the DP model decoherence is quantified by a ``self-interaction'' between the superposed amplitudes of a system: for a rigid body  (and in particular for an elementary mass) $\Gamma_{DP}=\frac{1}{2\hbar}[U(XX)+U(YY)-2U(XY)]$, where $U(XY)=-G\int d^3r\int d^3r'\frac{f_X(r)f_Y(r')}{|r-r'|}$ is a gravitational energy between two mass-distributions $f_X, f_Y$ which are here associated with two superposed configurations $X, Y$ of \textit{the same system}~\cite{Diosi2007NJP}. 
By contrast, in {the KTM approach} decoherence depends on the gravitational interaction between \textit{different systems}. As a result, both frameworks predict decoherence in the position basis, whose magnitude is related to gravity, but differ both quantitatively (Table~\ref{testtable}) and conceptually (Table~\ref{KTMvsDP}) as follows:
\renewcommand{\arraystretch}{1.2}
\begin{table*}[ht!]
\caption{General form of the decoherence rates $\Gdp$ (Diosi-Penrose) and  $\tGktm^{min}$ (Kafri-Taylor-Milburn)~\eqref{gamma_multipart_min}  for spherical mass distributions  \cite{Diosi2007NJP, Pikovski2009Master}.   $\delta$ denotes the cut-off of the DP model, and $\Delta x$ is the superposition size. Case (iii) considers decoherence of the CM of a body comprising $N_1$ constituents of mass $m$ in the presence of another body comprising $N_2$  masses $m$.  Whereas $\tGktm^{min}$ depends on the gravitational force gradients between \textit{different} particles, $\Gdp$  depends on the \textit{self-interaction} between superposed amplitudes of the same particle.  \hspace*{\fill}}
\centering
\begin{tabular}{||c l | c| c||}
\toprule
          & \textbf{Scenario:}        & ${\Gdp}$                            & $\tGktm^{min}$ \\ \hline
$(i)$ & \textbf{single particle} & $\frac{Gm^2\Delta x^2}{2\delta^3\hbar}$ for $\delta\gg\Delta x$ & 0 \\
        &   mass $m$    & $ \frac{2Gm^2}{\delta\hbar}\left(\frac{6}{5}-\frac{\delta}{\Delta x}\right)$ for $\delta\ll\Delta x$ & \\\hline

$(ii)$ &  \textbf{two particles}  &                     &   \\       
         &  masses $m, M$; distance $d$     & same as (i)  & $\frac{GmM\Delta x^2}{d^3\hbar}=\left\{
                \begin{array}{ll}
               0, \;d\rightarrow \infty \\
               \frac{mc^2}{\hbar}(\frac{\Delta x}{R_{S}})^2, \; d\rightarrow R_{S}:=\frac{2GM}{c^2}\\ 
               \end{array} 
               \right. $ \\ \hline        
$(iii)$ & \textbf{two composite bodies; }   &  $N_1\frac{Gm^2\Delta x^2}{2\delta^3\hbar}$; $\delta\gg\Delta x$       &                        \\ 
	  & masses $N_1m$, $N_2m$; dist.~$d_{ij}$     &  $N_1\frac{2Gm^2}{\delta\hbar}\left(\frac{6}{5}-\frac{\delta}{\Delta x}\right)$ for $\delta\ll\Delta x$&$\displaystyle \frac{ \Delta x^2}{2\hbar}\left(\sum_{i\neq j =1}^{N_1} |K_{ij}|+\sum_{i =1}^{N_1}\sum_{j=N_1+1}^{N_1+N_2}|K_{ij}|\right)$ \\  	 \hline\hline
% \bottomrule	  
\end{tabular}
\label{KTMvsDP}
\end{table*}
\paragraph{} For a point particle $\Gamma_{DP}$ diverges and requires a cut-off $\delta$ in the \textit{coherent spread} of the particle's wave-function \cite{Diosi2007NJP}, whereas the KTM {approach} is well-defined for point particles.
\paragraph{} A single elementary particle in an otherwise empty universe decoheres in the DP model, %(arbitrarily fast, for arbitrary small $\delta$), 
but does not  decohere in the KTM {{approach}}: if other systems are removed far from the particle  
 $\tilde\Gamma_{KTM}^{min}\rightarrow 0$, Table~\ref{KTMvsDP} row (i).
This is an important feature since for a single particle in an otherwise empty universe 
the notion of ``location'' has no physical meaning. Thus, arguing that the particle is -- or is not -- in 
a superposition of ``two different locations'' has no physical meaning either -- and the scenario cannot give rise to any physical effect.
\paragraph{}  KTM decoherence crucially depends on the distance $d$ between the test particle and other masses: For fixed $M$ and $\Delta x$: $0<\Gktm^{min}<\frac{mc^2}{\hbar}(\frac{\Delta x}{R_{S}})^2$ where the lower bound holds for $d\rightarrow\infty$ and the upper for $d=R_{S}=2GM/c^2$ (the Schwarzschild radius of $M$), Table~\ref{KTMvsDP} row (ii). In contrast, $\Gamma_{DP}$ for a single particle is independent of its gravitational environment. 
\paragraph{} %The KTM model requires a cut-off in the minimum distance between elementary masses.  Moreover, 
The KTM ``self interaction'' terms -- first sum in eq.~\eqref{decoh_min}, Table~\ref{KTMvsDP} row (iii) -- are purely classical: They connect \textit{different constituents} of a composite system, not different points of a single system wave-function.
\paragraph{} The {KTM proposal} predicts vanishing decoherence when all force gradients $K_{ij}$ are negligible, i.e.~% the linear terms in particles' positions can remain finite and the 
the sum of the homogeneous field contributions is induced without decoherence. {It is thus }compatible (to a limited extent) with the equivalence principle, as it does not predict any decoherence in the above case as well as for an accelerating particle.

\section{Discussion\;}

 We have shown that an LOCC gravity framework is very strongly constrained by experiments. Our analysis relies on certain auxiliary assumptions,  particularly  regarding the mass distribution of the earth and that all $N$ laser pulses comprising each $\pi$ and $\pi/2$ atom-light interaction are applied effectively simultaneously.  While these assumptions do not seem to pose a formidable challenge to our conclusions,  one certainly can improve on the  presented analysis: one example is to use an atom-fountain  gravity-gradiometer (two interferometers with vertical separation $L$) and a large mass $M$ (in the plane of, say, the lower interferometer), whose horizontal distance $d_h$ to the atoms can be varied \cite{rosi2014precision, PhysRevLett.114.013001, asenbaum2016phase}, cf.~Table~\ref{testtable}. A continuous mode of operation could be considered for improved sensitivity \cite{Geiger2017PRL}. 
The KTM  proposal predicts a different phase noise in the two interferometers as a function of $d_h$. With $M=252$ kg and $0.25<d_h<0.5$~m, an experiment at LMT order $10\hbar k$ and $T=0.5$ s would see the lower interferometer's contrast  varying between $0.5$---$0.65$, while the upper -- between $0.62$---$0.64$. %The setup of ref.~\cite{asenbaum2016phase} already used an  $84$ kg source mass -- our proposed experiment would use two additional such masses in this setup. Realization of such an experiment would allow relaxing our present assumptions about the mass distribution of the Earth.

%\pau{(Same concern here as before. I would probably say something like: To address any potential concerns regarding the analyzed experimental setup, it would be desirable ... )} 
%\tcr{To avoid potential systematic errors, } it would be desirable to conduct also non-interferometric tests of the \tcr{classical channel gravity}. %(e.g.~to avoid potential systematic errors). %affect the conclusions). 
 {Thus far, tests of the KTM framework were suggested with optomechanical or torsion balance setups. However, even including Earth into the analysis, as in the present work, such tests would face a formidable challenge.} For an optomechanical experiment, in order to detect KTM decoherence on top of the thermal noise, the mechanical frequency $\Omega$, quality factor $Q$ and the temperature $T$  of the mechanical oscillator must satisfy $ T\Omega/Q<G\hbar M_\oplus/2 k_B R_\oplus^3\sim10^{-18}$ K/s, where $k_B$ is the Boltzmann constant; a state of the art setup \cite{Schnabel:2015.PRA} with $Q=2\times10^7$, $\Omega/2\pi=1$ Hz, $T=4$ K yields $T\Omega/Q\sim10^{-6}$ K/s. For the {original KTM model} to be discernible from  measurement noise, the measurement frequency $\omega$  must satisfy $\omega^2<G M_\oplus/R_\oplus^3\sim10^{-6}$ Hz$^{2}$, whereas the value considered in \cite{Schnabel:2015.PRA} (at the standard quantum limit) gives $\omega^2\sim10^6$ Hz$^2$. Current optomechanical sensitivities would thus still need to be improved in order to test KTM assumptions.
For torsion balance setups with $1$---$10$ kg masses KTM approach yields $\Gktm^{min}~\sim 10^{25}$ Hz, 
 while experimental bound obtained from refs~\cite{PhysRevLett.111.101102, RevModPhys.88.035009}, 
 is $\sim 10^{40}$ Hz (see Appendix \ref{app:torsion_bal}).  

 From the theoretical-physics perspective, an immediate question is how much entanglement (what channel capacity) suffices to reproduce the experimental results? In order to address this question, it would be desirable to formulate an extension of the KTM approach -- describing gravitational interactions beyond the Newtonian limit, as well as allowing for larger channel capacities.   %the experimental constraints on the KTM approach mean that 
(Any ``complete" framework that has Newtonian gravity as its low-energy limit cannot be mediated by DOFs which are fundamentally classical and interact with the constituents of massive bodies in a local fashion -- as this would require the Newtonian limit of such a framework to be compatible with an LOCC channel, which contradicts experiments.) %Since classical channel gravity  essentially describes the dynamics of  an open quantum system, formulating its relativistic version might be more approachable than for other decoherence models -- there is still no covariant version of such well-studied models as continuous spontaneous localization \cite{CSLmodel, Bassi:2012bg}, or Diosi-Penrose. This expectation is supported by the fact that 
We note here that three models exploring various aspects of the KTM approach in different relativistic contexts have already been put forward \cite{Altamirano:2017EmergentDark, khosla2017detecting, pascalie2017emergent} and frameworks with an increased channel capacity, can be constructed %based on its quantum collisional model formulation -- since collisional models can give rise to arbitrary system dynamics, ranging from exact unitarity to arbitrarily fast decoherence~\cite{Altamirano2016knc}.  The required capacity and scaling with the system's mass can be achieved 
e.g.~by relaxing the assumption of local system-ancilla interactions or constraining the amount of energy introduced to the system by the ancilla  \cite{khosla2017detecting}. Empirically constraining such models and understanding their ramifications for the gravitationally induced entanglement remains an interesting subject for further investigation.

\section{Conclusion}

Results of the recent atom interference experiments very strongly constrain the worldview in which gravity reduces to the Newtonian pair potential at low energies \textit{and} is also fundamentally classical: mediated by LOCC channels %inducing Newtonian potential 
acting pairwise between atoms. %While existing experiments do not prove that gravity can entangle two masses, they can already refute the position that this is fundamentally not possible. \pau{Are we keeping the previous sentence? the referee did not like it.}  \tcb{I like this paragraph, I am wondering if we should move it to the first paragraph of the discussion section since it will have more exposure.}
%
%This is particularly noteworthy since testing gravity-mediated entanglement~\cite{marletto2017witness, marletto2017entanglement, bose2017spin} is still challenging.
%It should be stressed that the KTM model assumes that quantum theory applies without restriction to external degrees of freedom of any mass, but the gravitational interaction between pairs of masses is nevertheless compatible with a classical description. 
%
We have further shown that  -- contrary to current belief -- the KTM framework is not equivalent to the DP model.
It is noteworthy that the same experiments do not constrain also other   { alternative models
(gravity-related or not)} including continuous spontaneous localisation, or Schr{\"o}dinger-Newton theory \cite{CSLmodel, giulini2013gravitationally, Bassi:2012bg}. This raises a question about the notion of classicality of gravity in the DP and other models, and about their information-theoretic aspects. It is considered that such models allow for gravity to remain classical, but in the light of our analysis such claims need to be clarified:  if the gravitational DOFs are to remain classical and yield Newtonian gravity at low-energies,  some non-locality in their interactions with the masses  needs to be allowed. An example of the latter case is semi-classical gravity,  where the gravitational DOFs interact with the mean position of the source mass \cite{Kiefer:2014sfr}.

 The understanding of the classical vs quantum properties of gravity
is far from clear \cite{HallReginatto:2018nonclassgrav} and needs further work. However, the fact that the KTM framework can be empirically tested opens a novel route of investigation, one that focuses on a robust information-theoretic characterization of channels implied by alternative approaches to quantizing (or not quantizing) gravity.
From a broader perspective,  our work demonstrates that general frameworks as well as specific models for gravitational decoherence previously thought to be out of reach can be experimentally tested. 

 \section{Acknowledgments}
The authors thank S.~Basiri Esfahani,  W.~Bowen, F.~Costa, L.~Diosi, S.~Forstner, K.~Khosla, G.~Milburn, and A.~Tilloy for discussions. This work was supported in part by the Natural Sciences and Engineering Research Council of Canada, ARC Centre of Excellence for Engineered Quantum Systems grant no.~CE110001013 and the University of Queensland through UQ Fellowships grant no.~2016000089. Research at Perimeter Institute is supported by the Government of Canada through Industry Canada,  by the Province of Ontario through the Ministry of Research and Innovation. P.C-U. gratefully acknowledges funding from CONACYT.  N.A., P.C-U. and R.B.M are grateful for the hospitality of the University of Queensland where this work was initiated. M.Z.~acknowledges the traditional owners of the land on which the University of Queensland is situated, the Turrbal and Jagera people.

%%%%%%%%%%%%%%%%%%%%%%%%%%%%%%%%%%%%%%%%%%%%%%%%%%%%%%%%%%%%%%%%%%%%%%%%%%
\bibliographystyle{linksen}
\bibliography{DatabazeG}

\onecolumngrid

\appendix

\section{Essentials of  { Classical Channel Gravity}}
\label{app:KTM}

We describe here the basic features of {Classical Channel Gravity
as originally discussed by KTM}
~\cite{Kafri:2014zsa}. The key premise is that Newtonian gravity is a fundamentally classical interaction that cannot increase entanglement between any two systems.  This premise is applicable to any non-relativistic system, though the original proposal considered  a pair of harmonic oscillators for testing that idea.

This  can most easily be derived in the context of collisional dynamics.
A collisional model with a time-scale $\tau$ describes evolution of the state as $\rho_s(t+\tau)=\Tr_{\C{A}}\{\h U(\tau)(\rho_{s}\otimes\rho_{a}) \h U^\dagger(\tau)\}$, where  $\rho_s$ and $\rho_{a}$ are the {density matrices} of the system and the environment (ancillae), respectively, and $\h U(\tau)$ describes their joint unitary evolution; {the trace is over the ancillae.}

 The {original} KTM model as defined in ref.~\cite{Kafri:2014zsa} considers a quasi one-dimensional setting of two essentially point-like massive particles. At each step of the collisional dynamics the massive particles  interact with ancillas $a_1, a_2$ via two interaction terms:  the ``measurement'' interaction: $\hat x_1\otimes\hat p_{a_1}+ \hat x_2\otimes\hat p_{a_2}$ {(where ancilla obtain information about the positions of the particles)} and the ``feedback'' interaction  $K \hat x_1\otimes\hat x_{a_2}+K \hat x_2\otimes\hat x_{a_1}$ {(where ancilla induce a force on the particles depending on the information about the position of the other mass, acquired in the ``measurement'' step)}. Here $\hat x_i$, $i=1,2$ are the position operators of the $i^{th}$ mass, and $\hat x_{a_j}, \hat p_{a_j}$, $j=1,2$ are position and momentum operators of the $j^{th}$ ancilla. For the state of ancillas giving rise to finite effective dynamics (e.g Gaussian states with width $\sigma$) and in the continuous-interaction limit of $\tau \to 0$ 
the following master equation results \cite{PhysRevA.36.5543,WisemanMilburn:Book:2010, jacobs2014BookQuantum, Altamirano2016knc}:
$$ \dot{\rho}_s=-\frac{i}{\hbar}[\h H_0+K\h x_1 \h x_2, \rho_{s}]- \left(\frac{1}{4D}+\frac{K^2D}{4\hbar^2}\right)\sum_{i=1,2}[\h x_i, [\h x_i,\rho_{s}]],$$ where $D:=\lim_{\tau \to 0, \sigma \to \infty} \tau\sigma$ -- {this corresponds to a limit in which increasingly imprecise measurement of broad width $\sigma$ occur with increasing rapidity $\tau$ such that the product $\tau\sigma$ remains finite \cite{Altamirano2016knc}.}
{The form of the ``measurement'' and ``feedback'' terms is fixed by the assumptions of local system--ancilla interactions and the desired form of  the effective system--system interaction, up to the rescaling by $\alpha$ and $1/\alpha$ of the ``measurement''  and ``feedback'' interactions, respectively. Such a reparametrisation, however,  returns the same model, i.e.~ with the same minimal bound on the decoherence rate.} 
We note that ref.~\cite{Kafri:2014zsa} studied a special case with $K:=2\frac{Gm_1m_2}{d^3}$ and $\h H_0$ describing two harmonic oscillators with masses $m_1, m_2$, where the effective unitary term $K\h x_1 \h x_2$  could be interpreted as the normal mode splitting due to an effective Newtonian potential.
 
By adding to the ``feedback'' interaction local terms  $-\frac{d^2K}{2}\left(\frac{1}{2} +\frac{\h x_1+\h x_2}{d}+\frac{\h x_1^2+\h x_2^2}{d^2}\right)$ acting trivially on the ancilla,  the effective unitary term becomes
$$
V_0=-G\frac{m_1m_2}{d}\left(1-\frac{x_1+x_2}{d} + \left(\frac{x_1+x_2}{d}\right)^2\right)
$$
 which approximates Newtonian gravitational potential $-G\frac{m_1m_2}{|X_1-X_2|}$ up to second order in $x_1, x_2$; $X_1-X_2\equiv d+x_1+x_2$. This  extension recovers the homogeneous part of the potential in addition to the force gradients and yields
\beq
 \dot{\rho}_s\!=\!-\frac{i}{\hbar}[\h H_0+V_0, \rho_{s}] 
 -\bigg(\frac{1}{4D}+\frac{K^2D}{4\hbar^2}\bigg)\!\sum_{i=1}^2[\h x_i, [\h x_i,\rho_{s}]].\label{two_part_master_1_supp}
 \eeq
which is the main-text eq.~(1).%\eqref{twopart_master_1}.
It can be applied with any $H_0$, e.g.~that of  two free particles with masses $m_1, m_2$. In such a case the overall unitary term (first commutator) describes the standard dynamics of two massive particles interacting via Newtonian potential,  while the non-unitary terms ensure that the resulting channel is entanglement non increasing.

{We emphasize that there is no requirement that the two massive systems are in any specific state, such as Gaussian states. However, specific states can be employed to  provide an independent argument for a lower bound on decoherence in the KTM {approach}: the KTM lower bound coincides with the lower bound that guarantees that entanglement specifically between Gaussian states does not increase~\cite{Kafri:2014zsa}. The key feature of the dynamics described in eq.~\eqref{two_part_master_1_supp} -- that it is entanglement non-increasing -- is independent of how it is derived: from a classical stochastic model of the environment (ancillae) \cite{WisemanMilburn:Book:2010} or from a quantum collisional model \cite{Altamirano2016knc}. It is a generic feature of LOCC channels (realized via local interactions and communication of classical information) that they cannot increase entanglement.}

\section{KTM model for composite systems}
\label{app:composite_grav}

Here we extend the KTM {approach by constructing a model for} composite systems in three dimensions relevant for the matter wave experiments we analyze in the main text. We consider two systems $s_1$ and $s_2$  {respectively consisting of $N_1$ and $N_2$ elementary constituents -- chosen to be atoms -- with masses $m_i, \;{i=1,..., N_1+N_2}$.  Choosing atoms to be the basic constituents of a body ensures that  (unlike the case for subatomic particles)  the total mass of a body is equal to the sum of its individual constituents.  Our aim  is to describe the behaviour of the centres of mass of two objects in the KTM model, thereby allowing a more complete comparison between it and the Diosi-Penrose model.  A multi-particle  extension of the KTM model could also be used directly (though perhaps more cumbersomely) and the same final results would be obtained.
}

\begin{figure}[h]
\centering
\includegraphics[width=5cm]{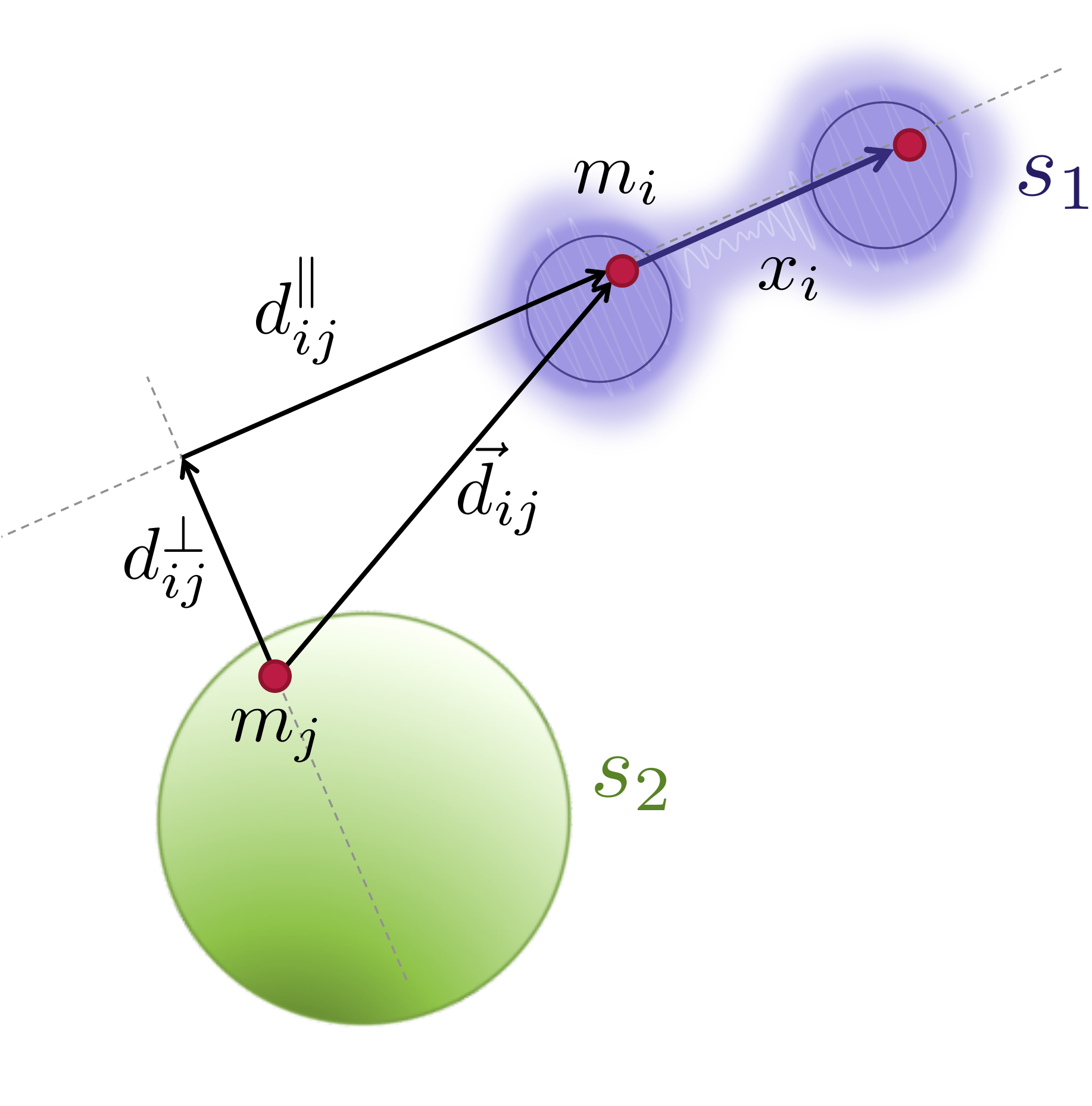}
\caption{Test body $s_1$ in a spatial superposition in the presence of a source mass $s_2$. For any pair of elementary masses $(m_i,m_j)$ forming the bodies, the distance $\vec d_{ij}$ between them can be decomposed into component $d_{ij}^\|$ along the direction of the spatial superposition of $s_1$ and a perpendicular component $d_{ij}^\bot$.  $x_i$   is the displacement from the initial position of the mass $m_i$, whose values span all locations between which the particle can be superposed. Note that the   assumption of rigidity implies that each constituent of $s_1$ is displaced by the same amount.}
\label{3DKTM}
\end{figure}

The classical gravitational potential energy between any two constituents $i, j$ reads 
$$
V_{ij} = \frac{Gm_i m_j}{|\vec r_{ij}|},
$$
where $\vec r_{ij}$ is the vector joining the positions of the individual masses $m_i, m_j$. We write this as
$\vec r_{ij}=\vec d_{ij}+\vec x_i+ \vec x_j$, where $\vec d_{ij}$ is the vector joining their
positions at the initial time and  $\vec x_{i,j}$ is the displacement of the CM of a given body. We consider the case where $s_1$ and $s_2$ are rigid. In applications of interest here,  $s_1$ will be a test mass (e.g.~an atom in an interferometer) and $s_2$ will describe matter gravitationally interacting with $s_1$,  (e.g.~the Earth).  We will thus assume that a) all constituents of a given body are in a superposition of equally distant locations (rigidity); b)  there is one distinguished direction defined by the superposition of the test mass (while the surrounding matter is well localised), see fig.~\ref{3DKTM}.

It is convenient to write  $\vec d_{ij}  = d_{ij}^{\|} \hat{e}+d_{ij}^{\bot}\hat{e}^{\bot}$ where $\vec x_i=x_i\hat e$ and thus  $\hat e$ is a unit vector  in the direction defined by the superposition and $\hat{e}^{\bot}$ is a unit vector in an orthogonal direction.  With the above we can write
\begin{eqnarray}\label{pot_expand}
V_{ij} &=& \frac{-Gm_i m_j}{\sqrt{(d_{ij}^{\|}+ x_i+x_j)^2+(d_{ij}^{\bot})^2 }}\approx  -Gm_i m_j\times\\
&\times &\left( \frac{1}{d_{ij}}-\frac{d_{ij}^{\|}}{d_{ij}^3}(x_i+x_j)+\frac{(d_{ij}^{\|})^2-\frac{1}{2}(d_{ij}^{\bot})^2}{d_{ij}^5}(x_i+x_j)^2\right)\nonumber
\end{eqnarray}
where ${d}_{ij} =\sqrt{(d_{ij}^{\|})^2+ (d_{ij}^{\bot})^2}$.

For any pair $(i, j)$ the ``measurement'' part of the interaction can thus be taken as  $\hat x_i\otimes\hat p_{m_{i}}+ \hat x_{j}\otimes\hat p_{m_{j}}$ and the ``feedback'' as $K_{ij} \hat x_{i}\otimes\hat x_{m_{j}}+K_{ij} \hat x_{j}\otimes\hat x_{m_{i}}+{\h Y_i\otimes\hat I_{m_j}+\h Y_j\otimes\hat I_{m_i}}$, where $\h Y_{i(j)}$ acts only on mass $m_{i(j)}$. The following master equation for the pair 
$$
{\dot\rho_{ij}}\approx-\frac{i}{\hbar}[\h H_0+\h Y_i+\h Y_j+K_{ij}\h x_i\h x_j, \rho_{ij}]-\Gamma_{ij}\left([\h x_i, [\h x_i,\rho_{ij}]]+[\h x_j, [\h x_j,\rho_{ij}]]\right)
$$
is thus obtained, where $\Gamma_{ij}\equiv\frac{1}{4D}+\frac{K_{ij}^{2}D}{4\hbar^2}$. 
Defining $K_{ij}:=2Gm_im_j\frac{(d_{ij}^{\|})^2-\frac{1}{2}(d_{ij}^{\bot})^2}{d_{ij}^5}$, $Y_i:=-Gm_i m_j( \frac{1}{2d_{ij}}-\frac{d_{ij}^{\|}}{d_{ij}^3}x_i+\frac{(d_{ij}^{\|})^2-\frac{1}{2}(d_{ij}^{\bot})^2}{d_{ij}^5}x_i^2)$ the master equation describes the induced (approximate) Newtonian interaction between $m_i$, $m_j$ (since $\h Y_i+\h Y_j+K_{ij}\h x_i\h x_j\approx V_{ij}$, eq.~\eqref{pot_expand})  with the decoherence term including a gradient of the Newtonian force between the pair, as in the eq.~\eqref{two_part_master_1_supp} but in three dimensions.
Note that the $x$-axis is defined by the direction of the superposition; that is why decoherence terms included in the model   also act only in this direction. 
The dynamics of all  $N_1+N_2$ constituents, described by  $\rho_{tot}$, reads
\begin{equation}
\label{many_body_master1}
\begin{aligned}
\dot\rho_{tot}&=-\frac{i}{\hbar}[\h H_0+\sum_{i< j}^{N_1+N_2} V_{ij} , \rho_{tot}]& 
 - \sum_{i < j}^{N_1+N_2}\Gamma_{ij}&\bigg([\h x_i, [\h x_i,\rho_{tot}]]+[\h x_j, [\h x_j, \rho_{tot}]]\bigg).
\end{aligned}
\end{equation}

\begin{figure}[h]
\centering
\includegraphics[width=10cm]{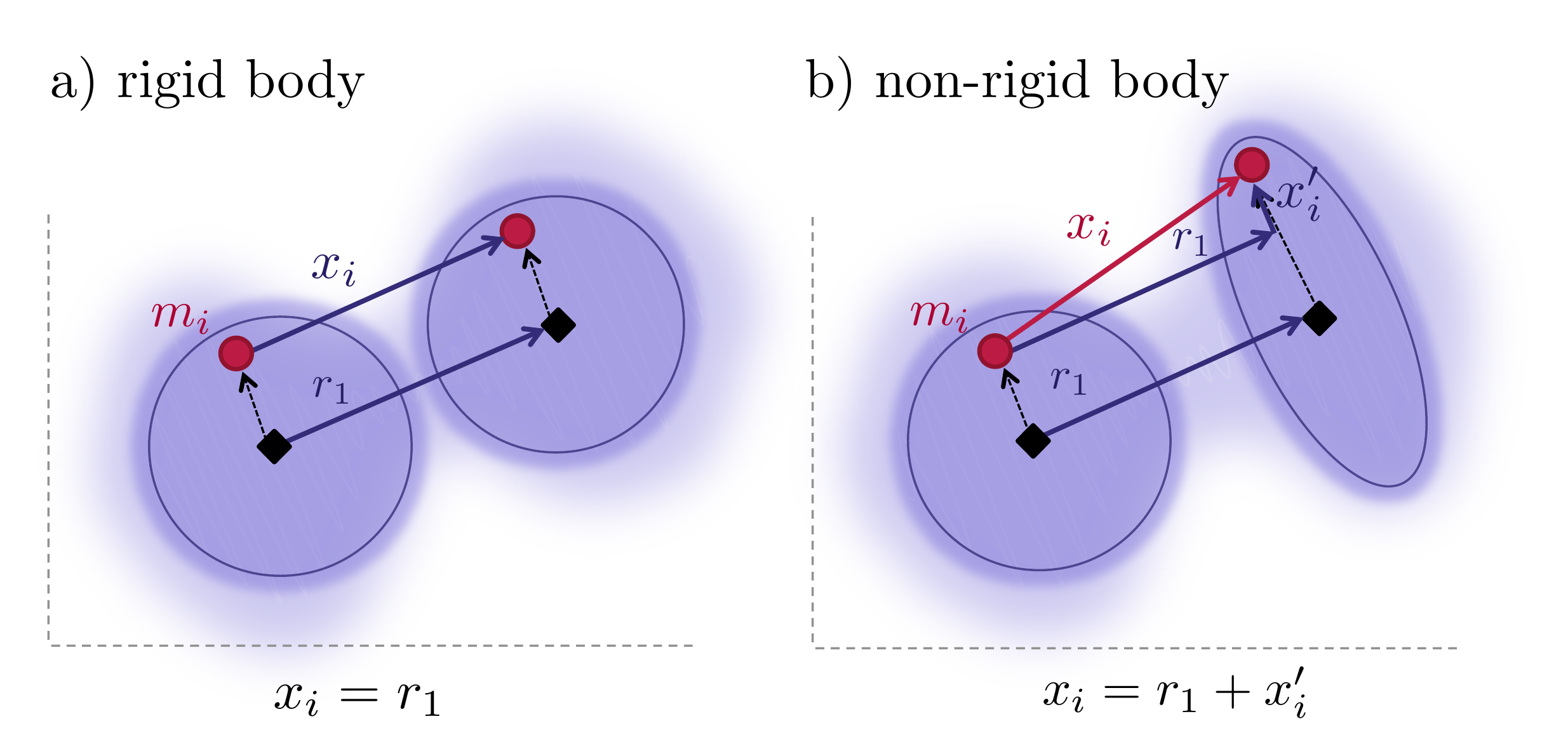}
\caption{Displacement $x_i$ of the $i^{th}$ constituent of a) rigid body, b) non-rigid body. For a rigid body each constituent remains at the same distance (dashed arrow) from from the centre mass (black diamond), and its displacement is the same as that of CM $x_i=r_1$. For a non-rigid body, the displacement of a constituent  can differ from that of the centre of mass, $x_i=r_1+x_i'$. This work  only considers case a).}
\label{relative_coord}
\end{figure}

Introducing  the displacement $r_1$  ($r_2$) of the CM of $s_1$ ($s_2$), and $\h x'_i$ as the displacement relative to the CM, the displacement of any individual constituent can be described by $\h x_i=\h r_1+\h x'_i$ for $i< N_1$ (for constituents of $s_1$) and $\h x_i=\h r_2+\h x'_i$ for $i>N_1$ (for constituents of $s_2$). 
With the above $[\h x_i, [\h x_i,\rho_{tot}]]=[\h r_1, [\h r_1,\rho_{tot}]]+[\h x'_i, [\h x'_i,\rho_{tot}]]+[\h r_1, [\h x'_i,\rho_{tot}]]+[\h x'_{i}, [\h r_1,\rho_{tot}]]$, for $i\leq N_1$ and analogously (with $\h r_2$ instead of $\h r_1$) for $i>N_1$. 
From the assumed rigidity of the bodies it follows that relative degrees of freedom are uncorrelated with the CM and their displacements remain negligible; see fig.~\ref{relative_coord}  for an illustration (i.e. a rigid body whose CM is in a superposition of locations $a$ and $b$ is described by a correlated state where all its constituents are at fixed distances relative to the CM position $a$ and at the same fixed distances relative to the CM position $b$).

Tracing over the relative degrees of freedom and keeping the CM positions of $s_1$ and $s_2$ results in the following master equation (in performing the trace, for simplicity one can assume that the CM of $s_i$ coincides with the position of one of its particles)  
\begin{equation}
\label{many_body_master2}
\begin{aligned}
\dot\rho_{s_{1}s_{2}}=& -\frac{i}{\hbar}[\h H_0+V , \rho_{s_{1}s_{2}}]   
 -2 \sum_{i< j=1}^{N_1}\Gamma_{ij}[\h r_1, [\h r_1,\rho_{s_1s_{2}}]] -  2\sum_{N_1<i< j}^{N_1+N_2}\Gamma_{ij} [\h r_{2}, [\h r_{2},\rho_{s_1s_{2}}]]  \\
&\qquad -\sum_{i=1}^{N_1}\sum_{j=N_1+1}^{N_1+N_2}\Gamma_{ij} \bigg([\h r_1, [\h r_1,\rho_{s_1s_{2}}]] + [\h r_{2} ,[\h r_{2},\rho_{s_1s_{2}}]]\bigg)
\end{aligned}
\end{equation}
where $V=\sum_{i< j}^{N_1+N_2} V_{ij}\approx -G\frac{M_1M_2}{|d+r_1+r_2|}$; $M_1$ and $M_2$ are correspondingly the total masses of $s_1$ and $s_2$. 

Finally, tracing over the degrees of freedom of $s_2$ one obtains the master equation for the CM of  $s_1$:
\begin{equation}
\label{composite_test_mass}
\begin{aligned}
\dot\rho_{s_{1}}&=-\frac{i}{\hbar}[\h H_0+V, \rho_{s_{1}}]  -\big(2 \sum_{i< j=1}^{N_1}\Gamma_{ij} +\sum_{i=1}^{N_1}\sum_{j=N_1+1}^{N_1+N_2}\Gamma_{ij}\big) [\h r_1, [\h r_1,\rho_{s_1}]],
\end{aligned}
\end{equation}
\begin{figure}[h]
\centering
\includegraphics[width=5cm]{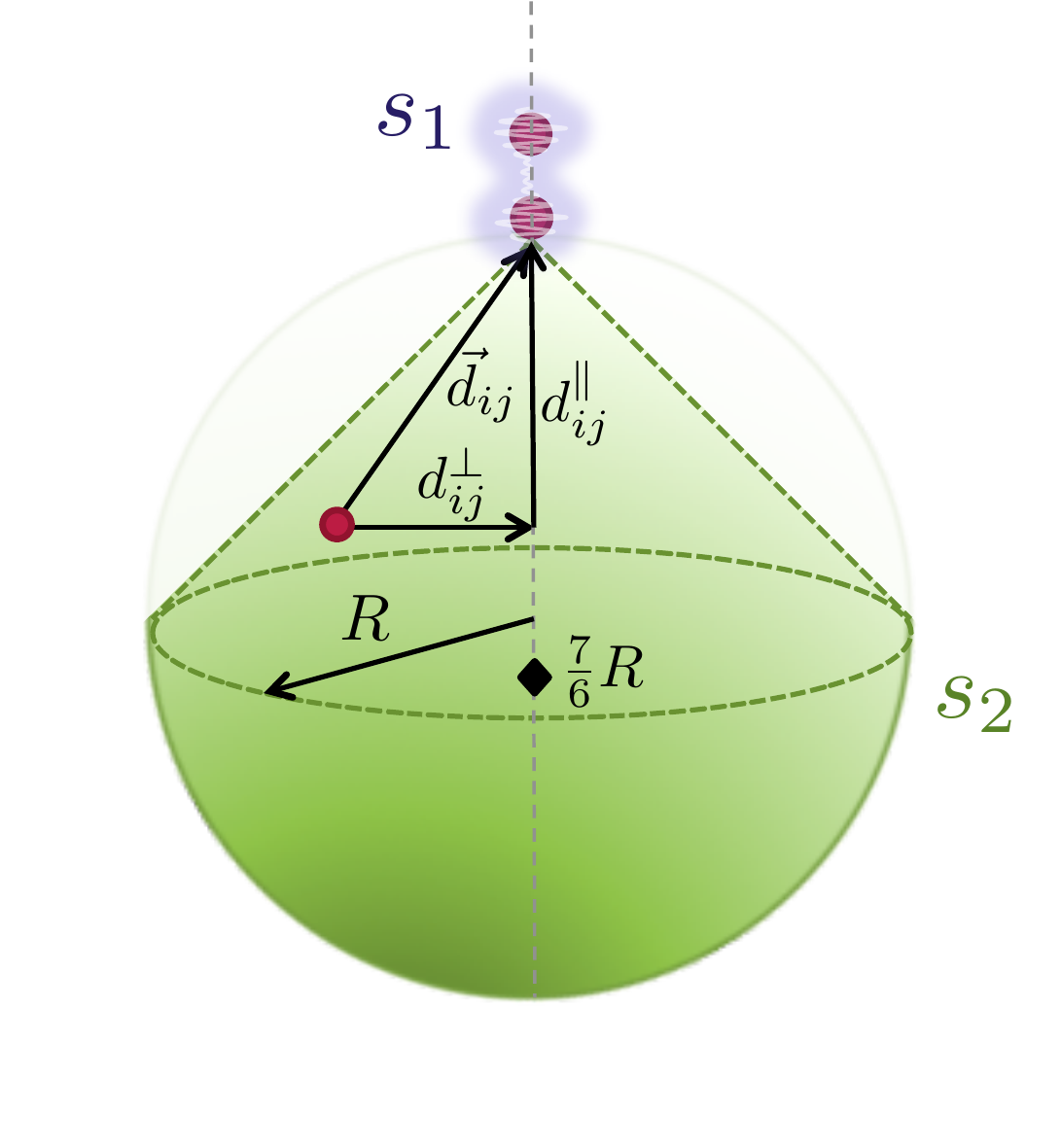}
\caption{Estimation of the decoherence effect on a single atom, $s_1$, due to Earth: for simplicity we consider only the portion of  Earth for which the decoherence rate is always greater than the effect stemming from matter concentrated at its centre of mass . This portion of Earth's mass is defined as all constituents (atoms) for which $d_{1j}^\bot<d_{1j}^\|$ -- inside the cone-and-half-ball, shaded green in the figure. The total mass of the region is 3/4 of the mass of Earth $M$, and its centre of mass is 7/6 of the Earth's radius $R$.  }
\label{convex}
\end{figure}

 We are particularly interested in the case when $s_1$ is a single atom, $N_1=1$, and $s_2$ is the entire Earth. Minimising the decoherence rate for  each pair $(1, j)$ gives $\Gamma_{1j}^{min}=\frac{K_{1j}}{2\hbar}$ and the total decoherence rate is given by $\sum_{j\in\mathrm{earth}}\frac{K_{1j}}{2\hbar}$. Note, that every constituent of the Earth acts so as to increase the decoherence rate of $s_1$. Here we seek to relate the resulting decoherence to that given by the Earth's CM, as in the original model. (For some geometries, the multi-particle formulation of the model, and the original KTM prediction for the CMs of the bodies give different results\footnote{Let $s_2$ to be a spherical shell of radius $r$ comprising N particles of mass $m$, and $s_1$ --  an elementary particle inside the shell. Sum over the constituents of the shell yields a finite decoherence rate, which can be made arbitrarily small by increasing $r\rightarrow\infty$. However, decoherence predicted by the model applied directly to the CM of $s_2$ is arbitrarily large for $s_1$ arbitrarily close to the shell's centre, independently of $r$.}.) Since $K_{1j}$ as a function of $(d_{1j}^{\|}, d_{1j}^\bot)$ is convex only for $|d_{1j}^{\|}|< |d_{1j}^\bot|/\sqrt{2}$ we consider only a portion $\mathcal{C}$ of the Earth's mass,  which lies within the volume where $K_{1j}$ is convex, see fig.~\ref{convex}.   

The
overall decoherence rate is greater than that stemming from particles in $\mathcal C$, which itself is greater than   decoherence coming from the centre of mass of $\mathcal C$. Since $|d_{1j}^{\|}|< |d_{1j}^\bot|/\sqrt{2}<|d_{1j}^\bot|$, the region $\mathcal C$ can be taken as a cone of height $R$ and support of area $\pi R^2$ together with a half ball of radius $R$. Assuming a constant density for body $M_2$, the mass of $\mathcal C$ is $\frac 3 4M_2$ and its CM is at a distance $\frac{7}{6}R$ from the top surface, as depicted in fig.~\ref{convex}. The quantity  
$$
\Gamma_{M_2;R}^{min}=\frac 3 4 \left(\frac 6 7\right)^3\Gamma_{KTM}(M_1, M_2, R) = 0.47\; \Gamma_{KTM}(M_1, M_2, R)
$$
where  $\Gamma_{KTM}(M_1, M_2, R)=\frac{GM_1M_2}{R^3}$ is the lower bound on the decoherence rate of mass $M_1$ due to the
presence of the homogeneous ball of mass $M_2$ and radius $R$. Note that in the above the $\Gamma_{KTM}(M_1, M_2, R)$   is the decoherence rate as per the original KTM model for elementary masses $M_1, M_2$ at a distance $R$.

{
Note that our overall result is lower bounded by the decoherence rate calculated as if the entire body were concentrated in the centre of mass.  Hence the predictions of the KTM  model for the experiments we analyze will not change if we choose  constituents of the Earth that are different from atoms.  }

{
We close this section by noting that in general there will be decoherence also in the transverse directions. 
However the setup we analyze is not sensitive to any loss of coherence in transverse directions. Furthermore, any such effects will  further increase decoherence rates, and thus will not affect our results. }

\section{Torsion Balance }
\label{app:torsion_bal}

 Here we consider an application of the KTM {approach} to torsion balance experiments. These experiments measure the  gravitational constant $G$ by  balancing the torque produced by the gravitational attraction between massive objects situated on the balance.  We show here how  the measurement/feedback model can be used to simulate a classical potential in this context.
\begin{figure}[h]
\centering
\includegraphics[width=5cm]{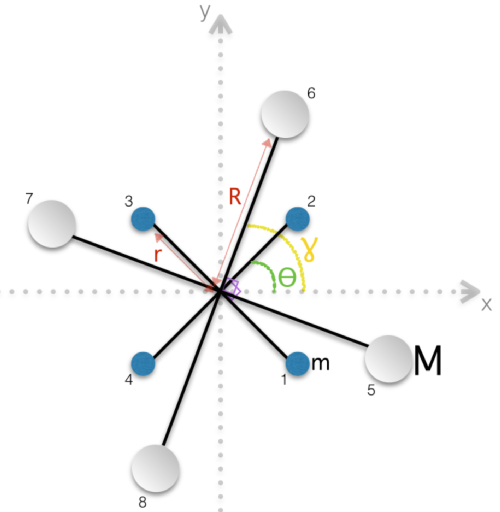}
\caption{Set-up of the torsion balance experiment. }
\label{tbal}
\end{figure}
 
The Hamiltonian of the system is
\begin{equation}
H=\sum_i \frac{1}{2}m_i\left(\dot{x}_i^2+\dot{y}_i^2\right)+\sum_{i,j}V_{ij}\,,
\label{H.torsion.balance}
\end{equation}
where $i$ runs over all (spherical) bodies in the experiment and $V_{ij}$ denotes the (Newtonian)
gravitational potential between pairs of bodies,  with each body regarded as a point-like
object located at its centre-of-mass. 

 The actual experiment consists of 8  bodies, 4 identical small bodies each of mass $m$, and 4 other identical large bodies each of mass $M$, as illustrated in figure \ref{tbal}.
Since all bodies in the experiment are in the same plane, we can write the above hamiltonian as 
\begin{align}
H&=2m\left(\dot{r}^2+r^2\dot{\theta}^2\right)+2M\left(\dot{R}^2+R^2\dot{\gamma}^2\right)+\sum_{i,j}V_{ij}\nonumber\\
&=2mr^2\dot{\theta}^2+2MR^2\dot{\gamma}^2+\sum_{i< j}V_{ij}
\end{align}
 in polar coordinates $(r,\theta)$ and $(R,\gamma)$ for the small and large bodies respectively, where the latter relation follows from the rigidity of the balance arms. The setup of the experiment ensures that $V_{ij}$ depends only on the variable $\alpha=\gamma-\theta$, and so
\begin{align}
H =&\frac{mr^2+MR^2}{4mr^2MR^2}p_\alpha^2+\frac{p_\xi^2}{4(mr^2+MR^2)}  
+\frac{4GmM}{\sqrt{r^2+R^2+2rR\cos\alpha}}+\frac{4GmM}{\sqrt{r^2+R^2+2rR\sin\alpha}}\nonumber\\
&+\frac{4GmM}{\sqrt{r^2+R^2-2rR\cos\alpha}}+\frac{4GmM}{\sqrt{r^2+R^2-2rR\sin\alpha}} +\frac{4Gm^2}{\sqrt{2}r}+\frac{4GM^2}{\sqrt{2}R}+\frac{Gm^2}{r}+\frac{GM^2}{R}
\label{H-ang}
\end{align}
 where $p_\alpha$ is the conjugate momentum to $\alpha$ and  $p_\xi$ is the conjugate momentum to the variable
$\xi \equiv \frac{mr^2\theta+MR^2\gamma}{mr^2+MR^2}$. 

 The relevant variable is small deviations of the angle $\alpha$ away from its equilibrium
 value $\alpha_0$.  Writing 
$\alpha=\alpha_0+\delta\alpha$, where $\delta\alpha\ll\alpha_0$,  the Hamiltonian can be approximated as
\be
H =  \frac{p_\alpha^2}{2 I_{eff}} +\frac{p_\xi^2}{4(mr^2+MR^2)}  +B\delta\alpha
+ C(\delta\alpha)^2 +\cdots
\ee
where $I_{eff} = \frac{2mr^2MR^2}{mr^2+MR^2}$ is the reduced moment of inertia  and
\begin{align}
B&=\sum_{n=1}^3\frac{4GmMrR \sin(\alpha_0+(n+1)\pi/2)}{\left(r^2+R^2-2rR\sin(\alpha_0+n\pi/2)\right)^{\frac{3}{2}}} \nonumber\\
C&= -\sum_{n=1}^3\frac{2GmMrR \sin(\alpha_0+n\pi/2)}{\left(r^2+R^2-2rR\sin(\alpha_0+n\pi/2)\right)^{\frac{3}{2}}}  +\sum_{n=1}^3\frac{6GmMr^2R^2 \sin^2(\alpha_0+(n+1)\pi/2)}{\left(r^2+R^2-2rR\sin(\alpha_0+n\pi/2)\right)^{\frac{5}{2}}} 
\end{align}
 and we have dropped the irrelevant constant terms from 
\eqref{H-ang}.

 In the context of the general master equation 
\eqref{two_part_master_1_supp}, we have  $\h x \to \delta \h\alpha \equiv \Delta \alpha$, and $K=2C$.  Writing $D = \frac{\hbar}{2C\epsilon}$, we obtain
\begin{align}
\dot{\rho}=-\frac{i}{\hbar}\left[ \frac{\h p_\alpha^2}{2 I_{eff}} +\frac{\h p_\xi^2}{4(mr^2+MR^2)}+B\delta \h\alpha+C\delta \h\alpha^2,\rho\right]- 
\frac{C}{2\hbar}\left(\epsilon+ \frac{1}{\epsilon} \right)\left[\delta \h\alpha,\left[\delta \h\alpha,\rho\right]\right]
\label{master-torsion}
\end{align}

To empirically constrain the KTM model using this class of experiments, we adopt the perspective that world measurements on Newton's 
constant $G$ \cite{RevModPhys.88.035009} have a certain degree of scatter, and that this scatter can provide a bound on how large the gravitational repeated-measurement-feedback effect from \eqref{master-torsion} is.  We have assumed that the bodies are pointlike, but as we have seen from earlier sections taking into account the finite size of the bodies will lead to corrections in $D$ (or $\Gamma$) that are of order unity.  
 
For the Cavendish experiment the total error in $G$ is
%\begin{widetext}
\begin{align}
\frac{\Delta G}{G} &= \frac{\delta(\Delta\alpha)}{\Delta\alpha} + 
\frac{\delta k}{k}  - \left( \frac{\delta M}{M} + \frac{\delta m}{m}  + 4 \frac{\delta r_a}{r_a} \right. 
 + \left.4 \frac{\delta r_b}{r_b}  - 5 \frac{\delta R_{ac}}{R_{ac}} - 5 \frac{\delta R_{b}}{R_{b}} + \alpha_{CT}
 \right) +   \frac{\delta \tau_c}{\tau_c}
\end{align}
where the meaning of these various quantities is explained in Ref. \cite{PhysRevLett.111.101102}
(see equation (11.10) therein). %\cite{PhysRevLett.111.101102} ). 
The minimal constraint on 
$\Delta\alpha$ is therefore
$$
\left| \frac{\delta(\Delta\alpha)}{\Delta\alpha} \right| \leq \left| \frac{\Delta G}{G}  \right|
$$
 where $\Delta\alpha = |\langle (\delta \h\alpha) \rangle|$ and
 $\delta(\Delta\alpha) = \sqrt{\langle (\delta \h\alpha)^2 \rangle}$, the latter quantity being the time-averaged variance computed from the repeated measurement process.  This quantity is 
$$
\langle (\delta \h\alpha)^2 \rangle = \frac{\hbar}{8 I_{eff}} T \left(\epsilon+ \frac{1}{\epsilon} \right)
$$
where $T$ is the timescale over which the experiment is performed.

To estimate the size of $\delta(\Delta\alpha)$, we have $m=1.2$ kg, $M=11$ kg, $r=120$ mm, $R=214$ mm, and $\alpha_0=18.9^{o}$, yielding $I_{eff} = 8.35\times 10^{-3}$  kg m$^2$.  
Hence
$$
\langle (\delta \h\alpha)^2 \rangle = 1.58 \times 10^{-33} T \left(\epsilon+ \frac{1}{\epsilon} \right) \textrm{s}^{-1}
$$
Consequently
\begin{eqnarray}
\left| \frac{\delta(\Delta\alpha)}{\Delta\alpha} \right| &=&\sqrt{1.58 \times 10^{-33} T \left(\epsilon+ \frac{1}{\epsilon} \right) }
 \leq \left| \frac{\Delta G}{G}  \right|  \sim 10^{-6} \nonumber  \\
 &&\Rightarrow T \left(\epsilon+ \frac{1}{\epsilon} \right) \leq \sim 6\times 10^{20}
\end{eqnarray}
 for ${\Delta\alpha} \sim 1$ radian.  For an experiment on the order of 1 day $=3600 \times 24 = 86400$ seconds, then
$$
\left(\epsilon+ \frac{1}{\epsilon} \right)\leq \sim 7 \times 10^{15}
$$
%So the parameter $\epsilon$ has two possible values
%\be
%\epsilon\leq\sim
%\begin{cases} 
%7\times 10^{15}\\
%1.4\times10^{-16}
%\end{cases}
%\ee
and the decoherence rate satisfies
 $$\frac{C}{2 \hbar}(\epsilon+\frac 1 \epsilon) \leq \sim 3.7\times10^{40}.
 $$
 
{A more sophisticated treatment involving the compositeness of the bodies in the torsion
balance will provide corrections of order unity to $C$. Since the above constraint is much weaker than
that provided by the fountain experiments we shall not pursue this any further.}

\end{document}